\documentstyle[prd,epsf,aps,floats]{revtex}

\def\Journal#1#2#3#4{{#1} {\bf #2}, #3 (#4)}

\def\PLB{Phys. Lett. B}
\def\PRL{Phys. Rev. Lett.}
\def\PRD{Phys. Rev. D}

\newcommand{\nn}{\nonumber}
\newcommand{\beq} {\begin{equation}}
\newcommand{\eeq} {\end{equation}}
\newcommand{\beqa} {\begin{eqnarray}}
\newcommand{\eeqa} {\end{eqnarray}}
\newcommand{\mrm}[1] {{\mathrm{#1}}}

\newcommand{\cf}{{\it cf.}}
\newcommand{\etal}{{\it et al.}}

\newcommand{\gev}{{\mrm{GeV}}}
\newcommand{\lqcd}{\Lambda_{QCD}}

\newcommand{\eq}[1]{(\ref{#1})}

\newcommand{\half}{$\scriptstyle \frac{1}{2}$ }

\def\lsim{\mathrel{\rlap{\lower4pt\hbox{\hskip1pt$\sim$}}
     \raise1pt\hbox{$<$}}}                
\def\gsim{\mathrel{\rlap{\lower4pt\hbox{\hskip1pt$\sim$}}
     \raise1pt\hbox{$>$}}}                

\begin{document}

\twocolumn[\hsize\textwidth\columnwidth\hsize\csname @twocolumnfalse\endcsname

\title{
\hbox to\hsize{\normalsize\hfil\rm NORDITA-2001-46 HE}
\hbox to\hsize{\normalsize\hfil\rm TTP01-28}
\hbox to\hsize{\normalsize\hfil hep-ph/0110297}
\hbox to\hsize{\normalsize\hfil 23 October 2001}
\vskip 40pt
Duality in Semi-Exclusive Processes}

\author{Patrik Ed\'en\footnotemark\ and Paul Hoyer\footnotemark\ }
\address{Nordita, Blegdamsvej 17, DK-2100 Copenhagen, Denmark}

\author{Alexander Khodjamirian\footnotemark\ }
\address{Institut f\"ur Theoretische Teilchenphysik, Universit\"at
  Karlsruhe, D-76128 Karlsruhe, Germany }

\maketitle

\vskip2.0pc
\begin{abstract}
Semi-exclusive processes like $\gamma p \to \pi^+ Y$ are closely analogous to
DIS, $ep \to eX$, in the limit where the momentum transfer $|t|$ to the pion
and the mass of the inclusive system $Y$ are large but still much smaller
than the total CM energy. We apply Bloom-Gilman duality
to this semi-exclusive process. The energy dependence of the $\gamma p \to
\pi^+ n$ cross section given by semi-local duality agrees with data for
moderate values of $|t|$, but its normalization is underestimated by about
two orders of magnitude. This indicates that rather high momentum transfers
are required for the validity of PQCD in the hard subprocess $\gamma u \to
\pi^+ d$. In the case of Compton scattering $\gamma p \to \gamma p$ the
analogous discrepancy is about one order of magnitude. In electroproduction
the virtuality of the incoming photon can be used to directly measure the
hardness of the subprocess.
\end{abstract}

\vskip2.0pc]

\footnotetext{$^*$Present address: Department of Theoretical Physics, Lund
University, Lund, Sweden.}
\footnotetext{$^\dag$On leave from the Department of Physics,
University of Helsinki, Finland.}
\footnotetext{$^\ddag$On leave from Yerevan Physics Institute, 375036
Yerevan, Armenia.}

The remarkable relation between Deep Inelastic Scattering (DIS) 
$eN \to eX $ and exclusive resonance production $eN \to eN^*$ known as
Bloom-Gilman duality \cite{BG} has recently been confirmed and extended by
data from JLab \cite{CEBAF}. Empirically,
\beqa\label{bgd}
\int\limits_{\delta x} d x F_2^{scaling}(x) &\simeq& F_2^{N^*}(Q^2) \nn\\
  &\propto& \frac{d\sigma}{dQ^2}(eN\to eN^*) \propto |F_{pN^*}(Q^2)|^2
\eeqa
where $F_2^{N^*}(Q^2)$ denotes the contribution of an $N^*$ resonance to the
DIS structure function $F_2$, and $F_{pN^*}(Q^2)$ is the 
exclusive $p \to N^*$ electromagnetic form factor. The Bjorken variable $x$
is related to the photon virtuality $Q^2$ and the invariant mass of the
hadronic system $W$ through
\beq\label{xbj}
x = \frac{Q^2}{W^2+Q^2-M_N^2}
\eeq
On the lhs of \eq{bgd}, the leading twist structure function
$F_2^{scaling}(x)$ is integrated over an interval $\delta x$ corresponding to
a fixed range of $W$ around the resonance mass, $W = M_{N^*}$. This semi-local
duality relation is approximately satisfied for each nucleon resonance region
including the Born term: $N^* = P_{11}(938),\ P_{33}(1232),\ S_{11}(1535)$ and
$F_{15}(1680)$ (where the regions are characterized by their main resonance).
The magnitude and $x$-dependence of the scaling structure function is thus
related to the magnitude and $Q^2$-dependence of the $N^*$ electromagnetic
form factors via the kinematic relation $x=x(Q^2,W^2=M_{N^*}^2)$ of Eq.
\eq{xbj}. Additional effects due to the logarithmic $Q^2$-dependence of
$F_2^{scaling}$ are discussed in Ref. \cite{cmlog}. The duality relation is
approximately satisfied even at low values of $Q^2$, for which one would not
expect perturbative QCD (PQCD) to apply.

The reasons for the validity of semi-local duality are something of a mystery
\cite{dgp,cec}. According to PQCD, the inclusive structure function is given
by incoherent scattering off single quarks in the target, whereas exclusive
form factors depend on a coherent sum over the quark charges. Interference
terms on the rhs of \eq{bgd} due to scattering on different target quarks
should systematically cancel only in an average over several resonances
\cite{ci}. Yet duality appears to be satisfied in a semi-local sense for both
proton and deuteron targets \cite{CEBAF}.

Duality in the $P_{33}(1232)$ region is not satisfied by the $\Delta$
resonance alone \cite{cmdelta}. The  electromagnetic form factor
$F_{p\Delta}(Q^2)$ falls faster with increasing $Q^2$ than predicted by the
dimensional scaling laws of PQCD \cite{ss}. The `background' under the
$P_{33}(1232)$ resonance becomes relatively more important with $Q^2$,
compensating for the resonance fall-off and maintaining semi-local duality in
the $P_{33}(1232)$ region. The corresponding phenomenon is not observed in
the $S_{11}(1535)$ and $F_{15}(1680)$ regions, where the resonance
contributions do obey dimensional scaling \cite{ss}. The spin dependence of
semi-local duality is discussed in Refs. \cite{cmspin,mel}.

\begin{figure}
\begin{center}
   \leavevmode
   \epsfxsize 0.45\textwidth
   \epsfbox{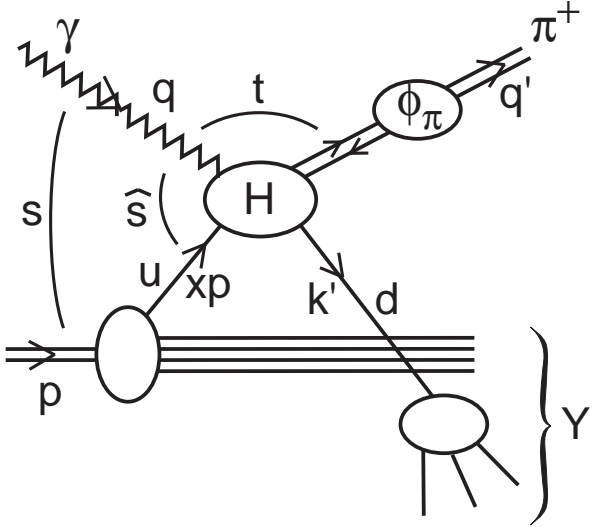}
\end{center}
\caption{{} Semi-exclusive scattering. In the limit $\lqcd^2 \ll |t|, M_Y^2
\ll s$ the cross section factorizes into a hard subprocess cross section
$\hat\sigma(H)$ times a target parton distribution. The subprocess $H$ acts
like a compact probe of target structure analogous to the virtual photon of
DIS.}
\end{figure}

In this paper we 
shall study semi-exclusive processes such as $\gamma p \to \pi^+ Y$ (Fig. 1).
In the kinematic limit where the total energy $s=(q+p)^2$ is much larger than
the mass of the inclusive system $Y$, 
$s \gg M_Y^2 \gg \lqcd^2$, the produced meson $\pi^+$ is separated from the
hadrons in $Y$ by a rapidity gap. When also the momentum transfer
$t=(q-q')^2$ is large the $\pi^+$ is produced in a compact configuration  
in the hard subprocess $H$ and, due to color transparency, does not
reinteract in the target. We have then a generalisation of ordinary DIS, with
the $eq \to eq$ subprocess replaced by $\gamma q \to \pi^+ q'$ and with the
physical cross section given by \cite{BDHP} 
\beq \label{sec}
\frac{d\sigma}{dx\,dt}(\gamma p \to \pi^+ Y) = \sum_{q,q'} q(x)
\frac{d\sigma}{dt}(\gamma q \to \pi^+q')
\eeq
where $q(x)$ is the distribution of the struck quark in the target.
From the point of view of the target physics, there is a one-to-one
correspondence between the semi-exclusive process and ordinary DIS, with $Q^2
\leftrightarrow -t$ and $W^2 \leftrightarrow M_Y^2$. The momentum fraction of
the struck quark is thus in the semi-exclusive process
\beq\label{xsm}
x = \frac{-t}{M_Y^2-t-M_N^2}
\eeq

The close analogy with DIS makes it natural to assume that semi-exclusive
processes obey Bloom-Gilman duality \cite{ACW} (see also \cite{DMS}).
Such duality links large momentum transfer exclusive cross sections to
standard DIS structure functions and, via Eq. \eq{bgd}, to exclusive form
factors. The multitude of processes that potentially can be compared in this
way may lead to a better understanding of the applicability of PQCD to
exclusive processes, and of Bloom-Gilman duality itself.

For the process of Fig. 1 semi-local duality predicts
\beq \label{sed}
\int_{\delta x}\left[u(x) + \bar{d}(x) \right] \frac{d\sigma}{dt}(\gamma u \to
\pi^+d) \simeq \frac{d\sigma}{dt}(\gamma p \to \pi^+ N^*)
\eeq
where the hard subprocess cross section is given by \cite{CW}
\beqa \label{subcross}
\frac{d\sigma}{dt}(\gamma u \to \pi^+d)&=& \frac{128 \pi^2
\alpha\alpha_s^2}{27 |t| \hat{s}^2}\left( \frac{e_u}{\hat{s}}+
\frac{e_d}{\hat{u}}\right)^2 (\hat{s}^2+\hat{u}^2) \nn\\
&\times&\left( \int_0^1\frac{dz}{z}\phi_\pi(z)\right)^2
\eeqa
Here $\hat s = (q+xp)^2 \simeq xs,\ \hat u \simeq -\hat s -t$ and
$\phi(z)$ is the pion distribution amplitude (normalized as $\int dz \phi(z) =
f_\pi/\sqrt{12}$ with $f_\pi=93$ MeV). In the semi-exclusive limit considered
in Ref. \cite{BDHP} $|t| \ll s$ and thus $\hat u \simeq -\hat s$. At large
$-t$ and fixed $M_Y^2 = M_{N^*}^2$ we have $x \to 1$ and thus $\hat s \simeq
s$.

The quark distributions are poorly measured at high $x$. Furthermore, target
mass corrections as well as the choice of $\delta x$ in \eq{sed} are sources
of uncertainty particularly for the nucleon Born term \cite{BG,mel,sim}. In
the present analysis we may, however, take advantage of the fact that the
target physics should be independent of the hard probe $H$ (\cf\ Fig. 1).
Hence the target mass corrections and $\delta x$ should be similar in Eqs.
\eq{bgd} and \eq{sed}. Noting also that the main contribution to the lhs is in
both cases from u-quarks allows us to eliminate the quark distributions in
favour of a direct relation between $\sigma(ep \to eN^*)$ and $\sigma(\gamma
p \to \pi^+ N^*)$,
\beq \label{dirrel}
\frac{d\sigma}{dt}(\gamma p \to \pi^+ n) =
16\pi^2\frac{\alpha\alpha_s^2 f_\pi^2}{|t|s^2}
\frac{G_{Mp}^2(-t)}{(1+r_{du}/4)}
\eeq
where we took the $s \gg |t|$ limit \cite{BDHP} in \eq{subcross} and assumed
the asymptotic form of the pion distribution amplitude, $\phi_\pi^{as}(z) =
\sqrt{3}f_\pi z(1-z)$. The ratio $r_{du}$ is the $d$-quark distribution
integrated over the interval $\delta x$, divided by the corresponding
quantity for the $u$-quark (we neglect the antiquark contributions since $x
\to 1$ at large $|t|$).

The $1/s^2$ energy dependence of the cross section \eq{dirrel} at fixed $|t|$
is a general consequence of the exchange of two spin \half quarks in the
$t$-channel. Higher order ladder corrections could modify the power. 

Data on $\gamma p \to \pi^+ N^*$ at large momentum transfer exists for
$E_\gamma \leq 7.5\ \gev$ \cite{Anderson}. The related measurements of the
semi-exclusive process $\gamma p \to \pi^+ Y$ were apparently never
published.

The measured $\gamma p \to \pi^+ n$ cross section is $\propto 1/s^2$ for $|t|
\lsim 2\ \gev^2$ \cite{Anderson}, in agreement with \eq{dirrel}. This
indicates the $t$-range for which the asymptotic energy dependence has set in
when $s \lsim 15\ \gev^2$. At $E_\gamma=7.5\ \gev$ and $t=-1.95\ \gev^2$ the
differential cross section is $3.45 \pm 0.16$ nb/GeV$^2$. Using
$G_{Mp}(Q^2)=2.79/(1+Q^2/.71\ \gev^2)^2$ \cite{ss}, $\alpha_s(2\ \gev^2)
\simeq 0.4$ \cite{Bethke} and $r_{du}=0.5$ the rhs. of Eq. \eq{dirrel} is
$0.065$ nb/GeV$^2$. Data is thus a factor $\sim 50$ larger than the duality
prediction!

At higher values of $|t|$ the $\gamma p \to \pi^+ n$ cross section
falls rapidly with energy, approximately as $1/s^6$ in the $|t| = 3\ldots 4\
\gev^2$ range \cite{Anderson}. Hence the duality relation
\eq{dirrel} will be better satisfied at larger $s$ and $|t|$. E.g.,
for $data/theory\ \propto 1/s^4$ their ratio approaches unity around
$E_\gamma \simeq 20$ GeV.

For DIS the duality relation \eq{bgd} works down to quite low values of $Q^2
\lsim 1\ \gev^2$ \cite{BG,CEBAF}. It is perhaps not surprising that the
subprocess $H$ in Fig. 1 requires higher values of $|t|$ to become compact --
the virtualities of the internal lines in the $\gamma u \to \pi^+ d$
subprocess are generally less than $|t|$. It is possible that duality
nevertheless applies to the semi-exclusive process $\gamma p \to \pi^+ Y$ at
lower values of $|t|$ in the sense that the shape of the scaling curve
measured at large $M_Y^2$ agrees on average with that of the resonance region.
Unfortunately, data on the semi-exclusive cross section is not available.

So far we discussed duality for $|t| \ll s$, a limit which appears
to be required for factorization in semi-exclusive processes \cite{BDHP}. 
The authors of Ref. \cite{ACW} took $|t| \sim s$
in the semi-exclusive process \cite{CW} and were led via duality
to a fixed angle limit of the exclusive process. Since $G_{Mp}^2(-t) \propto
1/t^4$, Eq. \eq{dirrel} evaluated at fixed $t/s$ predicts that
$d\sigma/dt(\gamma p \to \pi^+ n) \propto 1/s^7$ at fixed angle \cite{ACW}.
This energy dependence agrees with the data which at $\theta_{CM}=90^\circ$
scales as $s^7 d\sigma/dt \simeq 8.3 \mu$b GeV$^{12}$ over practically the
whole measured energy range \cite{Anderson}. 

However, Eq. \eq{dirrel} (with a
slight change of normalization due to evaluating \eq{subcross} at fixed angle)
underestimates the measured cross section by more than a factor 200. This
disagreement will not improve at higher energies since both data and theory
obey the same scaling. We conclude that duality, and more generally the
factorization of semi-exclusive processes implied by Fig. 1 and Eq.
\eq{sec},  can be valid only in the $|t| \ll s$ limit \cite{BDHP}. In the
fixed angle limit diagrams involving more than one target parton can
apparently not be neglected. This was also observed in Ref. \cite{HK}.

In the case of the process $\gamma p \to \pi^0 p$ we note that
the leading order subprocess cross section \eq{subcross} vanishes for $|t| \ll
s$. The cross section measured for this process \cite{Anderson,Shupe} is
actually somewhat larger than that for $\gamma p \to\pi^+ n$. This 
confirms our conclusion above that the expression \eq{subcross} is not
applicable  for the present data.

For Compton scattering, $\gamma p \to \gamma p$, the duality prediction
analogous to \eq{dirrel} is
\beq \label{compton}
\frac{d\sigma}{dt}(\gamma p \to \gamma p) = \frac{16\pi \alpha^2}{9s^2}
\frac{1+r_{du}/16}{1+r_{du}/4} G_{Mp}^2(-t) 
\eeq
With the same parameters as above this gives a Compton cross section
which is a factor $\sim 13$ below the data \cite{Shupe} at $E_\gamma = 6$ GeV,
$-t=2.45\ \gev^2$. The discrepancy is thus less than for $\gamma p
\to \pi^+ n$ but still substantial. The data indicates a $\sim 1/s^3$ energy
dependence of the cross section, again suggesting that higher values of both
$s$ and $|t|$ are likely to be required for agreement with PQCD.

To conclude, we applied Bloom-Gilman duality to semi-exclusive processes. In
a fixed angle limit \cite{ACW}, duality predicts $d\sigma/dt(\gamma p \to
\pi^+ n) \propto 1/s^7$ in agreement with data \cite{Anderson}. However, the
normalization of the $90^\circ$ data is more than two orders of magnitude
larger than expected from duality. Due to the identical scaling of both data
and theory this discrepancy will not decrease with energy. It thus appears
that the dynamics of semi-exclusive processes does not factorize into a a
hard subprocess and a target parton distribution (\cf\ Fig. 1) in a fixed
angle limit.

The energy dependence of the $\gamma p \to \pi^+ n$ cross section is
consistent with $1/s^2$ at fixed $|t| \lsim 2\ \gev^2$ as required by duality.
However, the duality relation underestimates the normalization by a factor 50
even at $|t| \simeq 2\ \gev^2$. At higher values of $|t|$ the data decreases
rapidly with energy, indicating that better agreement with the normalization
given by the subprocess cross section \eq{subcross} will be obtained at
higher values of $s$ and $|t|$. Such a conclusion is consistent with
preliminary data on semi-exclusive $\rho$ photoproduction at HERA, $\gamma p
\to \rho Y$, which suggests \cite{Crittenden} that large values of $|t| \gsim
8\ \gev^2$ are required for agreement with PQCD predictions of the dominance
of longitudinal $\rho$ polarization.

The empirical success of duality down to $Q^2 \lsim 1\ \gev^2$ in DIS
suggests that the shape of the
scaling curve may average the resonance contributions at
similar values of $|t|$ also in semi-exclusive processes. It would thus be
important to measure the $\gamma p \to \pi^+ Y$ cross section at higher
$M_Y^2$ (but consistent with the condition $M_Y^2 \ll s$). This should be
feasible at HERMES, COMPASS and HERA.

The effective size of the hard probe $H$ in Fig. 1 can be `measured' via its
dependence on the virtuality $Q^2$ of the initial photon. The subprocess cross
section will be independent of $Q^2$ as long as the transverse size of $H$ is
$\lsim 1/Q$. Conversely, the probe $H$ can be made more compact by increasing
$Q^2$. This should improve the agreement with the PQCD cross section
\eq{subcross} (modified to account for the photon virtuality).

\section*{Acknowledgments}
It is a pleasure to acknowledge helpful discussions with Stan Brodsky, Carl
Carlson, Markus Diehl, Jean-Marc Laget, Stephane Peign\'e and Mark Strikman.
This work has been supported in part by the European Commission under contract
HPRN-CT-2000-00130 (PE and PH). A.K. acknowledges the support
of STINT (Swedish Foundation for International
Cooperation in Research and Higher Education) during his visit to the
Department of Theoretical Physics, Lund University, and the support 
of BMBF (Bundesministerium f\"ur Bildung und Forschung).


\begin{thebibliography}{99}

\bibitem{BG}
E.~D.~Bloom and F.~J.~Gilman, \Journal{\PRL}{25}{1140}{1970} and
\Journal{\PRD}{4}{2901}{1971}.

\bibitem{CEBAF}
I.~Niculescu \etal, \Journal{\PRL}{85}{1182 and 1186}{2000}.

\bibitem{cmlog} C. E. Carlson and N. C. Mukhopadhyay,
\Journal{\PRL}{74}{1288}{1995} [hep-ph/9410351].

\bibitem{dgp}
A. de Rujula, H. Georgi and H. D. Politzer, \Journal{Ann.
Phys.}{103}{315}{1977}.

\bibitem{cec}
C. E. Carlson and N. C. Mukhopadhyay, \Journal{\PRD}{41}{2343}{1990};\\
C. E. Carlson, hep-ph/0005169.

\bibitem{ci}
F. E. Close and N. Isgur, \Journal{\PLB}{509}{81}{2001} [hep-ph/0102067].

\bibitem{cmdelta}
C. E. Carlson and N. C. Mukhopadhyay, \Journal{\PRD}{47}{1737}{1993} and
\Journal{\PRL}{81}{2646}{1998} [hep-ph/9804356].

\bibitem{ss}
G. Sterman and P. Stoler, \Journal{Ann. Rev. Nucl. Part. Sci.}{47}{193}{1997}
[hep-ph/9708370].

\bibitem{cmspin}
C. E. Carlson and N. C. Mukhopadhyay, \Journal{\PRD}{58}{094029}{1998}.

\bibitem{mel} W. Melnitchouk, \Journal{\PRL}{86}{35}{2001} [hep-ph/0106073].

\bibitem{BDHP}
S.~J.~Brodsky, M.~Diehl, P.~Hoyer and S.~Peigne,
\Journal{\PLB}{449}{306}{1999} [hep-ph/9812277].

\bibitem{ACW}
A.~Afanasev, C.~E.~Carlson and C.~Wahlquist,
\Journal{\PRD}{62}{074011}{2000} [hep-ph/0002271].

\bibitem{DMS}
R. Blankenbecler and S. J. Brodsky, \Journal{\PRD}{10}{2973}{1974};\\
 D. M. Scott, \Journal{\PRD}{10}{3117}{1974} and
\Journal{\PLB}{59}{171}{1975}.

\bibitem{CW}
C.~E.~Carlson and A.~B.~Wakely,
\Journal{\PRD}{48}{2000}{1993};\\
A.~Afanasev, C.~E.~Carlson and C.~Wahlquist,
\Journal{\PLB}{398}{393}{1997} [hep-ph/9701215] and
\Journal{\PRD}{58}{054007}{1998} [hep-ph/9706522].

\bibitem{sim} S. Simula, \Journal{\PLB}{481}{14}{2000} [hep-ph/9912067].

\bibitem{Anderson}
R.~L.~Anderson {\it et al.}, \Journal{\PRL}{30}{627}{1973};
\Journal{\PRD}{14}{679}{1976}.

\bibitem{Bethke} S. Bethke, \Journal{J. Phys. G}{26}{R27}{2000}
[hep-ex/0004021].

\bibitem{HK} H. W. Huang and P. Kroll, Eur. Phys. J. C {\bf 17}, 423 (2000)
[hep-ph/0005318].

\bibitem{Shupe} M. A. Shupe \etal, \Journal{\PRD}{19}{1921}{1979}.

\bibitem{Crittenden} J. A. Crittenden, in {\em Proceedings of DPF2000: The
meeting of the Division of Particles and Fields.} American Physical Society,
Columbus, Ohio, USA, 2000, hep-ex/0010079;\\
ZEUS Collaboration, Paper \#556 submitted to the Int. Europhys. Conf. on High
Energy Physics, Budapest, Hungary (July 2001).

\end{thebibliography}
\end{document}